\documentclass[aps,prl,10pt, twocolumn, superscriptaddress, amsmath, amssymb,shortbibliography]{revtex4-2}
\usepackage[british]{babel}

\usepackage{graphicx}   
\usepackage{bm}          
\usepackage{verbatim}  
\usepackage{times}

\usepackage{mathtools}
\usepackage{physics} 
\usepackage{siunitx}

\usepackage[dvipsnames]{xcolor}
\definecolor{mygreen}{RGB}{44,85,17}
\definecolor{myblue}{RGB}{34,31,150}
\definecolor{myred}{RGB}{255,66,56}

\usepackage[breaklinks=true,colorlinks=true,linkcolor=myblue,urlcolor=myblue,citecolor=myblue]{hyperref}

\begin{document}

% Title, authors, research centres and date

\title{Global Quantum Thermometry}
\author{Jes\'{u}s Rubio}
\email{J.Rubio-Jimenez@exeter.ac.uk}
\affiliation{Department of Physics and Astronomy, University of Exeter, Stocker Road, Exeter EX4 4QL, United Kingdom.}
\author{Janet Anders}
\affiliation{Department of Physics and Astronomy, University of Exeter, Stocker Road, Exeter EX4 4QL, United Kingdom.}
\affiliation{Institut f\"ur Physik und Astronomie, University of Potsdam, 14476 Potsdam, Germany.}
\author{Luis A. Correa}
\affiliation{Department of Physics and Astronomy, University of Exeter, Stocker Road, Exeter EX4 4QL, United Kingdom.}

\date{\today}
   
% Abstract

\begin{abstract}

A paradigm shift in quantum thermometry is proposed. To date, thermometry has relied on local estimation, which is useful to reduce statistical fluctuations once the temperature is very well known. In order to estimate temperatures in cases where few measurement data or no substantial prior knowledge are available, we build instead a theory of global quantum thermometry. Based on scaling arguments, a mean logarithmic error is shown here to be the correct figure of merit for thermometry. Its full minimisation provides an operational and optimal rule to post-process measurements into a temperature reading, and it establishes a global precision limit. We apply these results to the simulated outcomes of measurements on a spin gas, finding that the local approach can lead to biased temperature estimates in cases where the global estimator converges to the true temperature. The global framework thus enables a reliable approach to data analysis in thermometry experiments.

\end{abstract}

\maketitle

% Body of the document

%--------------------------------------

Quantum thermometry aims to improve precision standards for temperature sensing in the quantum regime \cite{stace2010quantum,DePasquale2018,mehboudi2019}. It can inform the design of nanoscale probes \cite{correa2015,plodzie2018fermion}, the choice of measurement \cite{mehboudi2015thermometry,correa2017,potts2019,mitchison2020fermi} and, as we shall see, the post-processing of measured data into an optimal temperature reading. Precision thermometry is rooted in the old problem of interpreting temperature fluctuations---in practice, temperature cannot be accessed directly but rather, estimated from the statistics of observable properties \cite{phillies1984, prosper1993, jahnke2011, falcioni2011, hemmen2013}. More generally, estimation theory allows to devise feasible strategies that can approach the fundamental precision limits of thermometry. Improving thermometric accuracy is not only relevant for quantum thermodynamics \cite{DePasquale2018}, but also in any practical application in which precise temperature control is necessary. 

Strategies for temperature estimation may be classified as follows. Let $T$ denote a true but unknown temperature, and let $\theta\in[\theta_1, \theta_2]$ represent a hypothesis about the value of $T$. Local estimation schemes assume that $\theta_2/\theta_1 \sim 1$. If no such \emph{a priori} hypothesis is required, the scheme is said to be global \footnote{`Local' and `global' are here defined with respect to the space of hypotheses for the true temperature, as is customary in quantum estimation theory \cite{paris2009}. However, the same terminology has been given different meaning when probes are multipartite: states and measurements associated with a part of the whole are said to be local, and global otherwise. These notions of `local' and `global' can also be found in quantum thermometry \cite{DePasquale2018, campbell2017}. However, the version of estimation theory used is still local in the standard sense, i.e., with respect to the parameter space. Therefore, the spatial partitioning of the probe is not relevant to the present discussion.}. One readily sees that, if $\theta_2/\theta_1 \sim 1$ and $T\in[\theta_1, \theta_2]$, then $\theta \sim T$. Consequently, local strategies allow to reduce statistical fluctuations once the temperature is well known \cite{rafal2015}, but they cannot address the estimation of unknown temperatures in full generality. Currently, most literature on quantum thermometry focuses on local protocols \cite{stace2010quantum,DePasquale2018,mehboudi2019,correa2015,plodzie2018fermion,mehboudi2015thermometry,correa2017,potts2019,mitchison2020fermi}. 

This is partly due to the widespread use of the Cram\'er--Rao bound (CRB) \cite{cramer1999mathematical,rao1992information} as the precision benchmark. The standard CRB assumes unbiased estimators, i.e., that the temperature estimates average to $T$, and it is exactly saturable only for a special class of probability models---the exponential family \cite{kay1993, rafal2015}. To accommodate a wider model set, one can employ local unbiased estimators \cite{fraser1964} which are appropriate when the unknown temperature lies initially on a very narrow interval \cite{rafal2015}. More generally, the CRB is approached using asymptotically large data sets \cite{kay1993, jesus2018}, which in turn reduces the estimation error down to a regime in which local strategies become optimal. The applicability of the CRB thus leads to schemes that are useful only in a local sense, excluding cases where little is known about the temperature \textit{a priori}, or where only few measurements can be performed. Furthermore, even if an exact saturation of the CRB were possible, the bound often depends explicitly on the unknown $T$ \cite{mehboudi2019}. In general, local thermometry is thus far too restrictive. 

This Letter puts forward a new theory for global quantum thermometry, that is, a theory applicable to estimates based on small data sets even if $\theta$ lies on a broad range. Here the problem of global temperature estimation is formulated, and fully solved, within the Bayesian framework \cite{jaynes2003, toussaint2011, jesus2019thesis}. We achieve this by assigning a prior probability reflecting the initial state of information about temperature, and identifying an appropriate measure of uncertainty (an averaged `cost' or `deviation'). The latter must also respect the scale invariance of the problem, and turns out to be a type of mean logarithmic error. Equipped with this measure of uncertainty, we are able to derive analytical expressions for the optimal temperature estimator and its uncertainty, both of which make no `unbiasedness' assumptions. Under certain conditions, local thermometry is recovered as a special case of this global formalism.

As a means of illustration, we apply the global theory to a non-interacting gas of $ n $ spin-$1/2$ particles. For this example, local thermometry is found unable to predict the true precision scaling when $n \lesssim 10^7$ (with $5\%$ tolerance). Moreover, the estimator identified as optimal in the local regime can in principle yield a much larger uncertainty than its global counterpart. To demonstrate the potential usefulness of the global approach in the analysis of experimental data, we also simulate and post-process measurement outcomes for this $n$-spin gas. The global estimator is then found to converge to the true temperature after $\mu \simeq 10^2$ trials. In contrast, a local analysis can lead to a biased temperature estimate even for large $\mu$. These results show that a paradigm shift towards Bayesian techniques may allow a more robust and significantly enhanced optimisation of thermometric protocols, especially in cases where the data are limited \cite{jesus2018, jesus2020mar, morelli2021}.

\vspace{0.5em}
\noindent\emph{\textbf{Scale invariance and logarithmic error.---}} Consider a system in equilibrium, where the true temperature $T$ is well defined but unknown. To infer it, one can perform energy measurements, which return the value $ E $, given $ T $, according to some conditional probability distribution. This plays the role of a likelihood function \cite{kay1993, jaynes2003}, as it links the measurement process with the unknown parameter. We denote such function by $ p(E\vert \theta) $, where we recall that $ \theta $ is our hypothesis for the true temperature $ T $. Instead of assuming that $\theta \sim T$, as local thermometry does, we introduce a prior probability $ p(\theta) $ as a proxy for our initial state of information about $ T $. It is then instructive to adopt the limit of complete ignorance, opposite to (and more general than) local estimation.

Naively, one would represent complete ignorance as $p(\theta) \propto 1$. However, the conditional probability $p(E\vert \theta)$ only depends on $\theta$ through the dimensionless ratio $ E/(k_B \theta)$, i.e.,
\begin{equation}
p(E \vert \theta)dE =  \frac{f \left[E/(k_B \theta)\right]}{\int d\hat{E} f [\hat{E}/(k_B \theta)]}dE
\label{model}
\end{equation}
for some function $f$ and where $k_B$ is Boltzmann's constant. This implies that we are dealing with a scale estimation problem \cite{jaynes2003, kass1996, toussaint2011}. Given that $T$ is, at this stage, completely unknown, so is the scale of the problem. Consistency thus demands our initial state of information to be invariant under transformations $ E \mapsto E' = \gamma E $ and $ \theta \mapsto \theta'= \gamma \theta $ \cite{jaynes2003, prosper1993}, where $ \gamma $ is a dimensionless constant. In turn, this means that our prior must satisfy $p(\theta) d\theta = p(\theta') d\theta'$, which leads to the functional equation $p(\theta) = \gamma\, p(\gamma \theta)$ with solution $p(\theta) \propto 1/\theta$ \cite{jaynes2003, toussaint2011}. Indeed, note that the problem could have been equivalently formulated in terms of the inverse temperature $\beta = 1/k_B \theta$. Such choice should not alter our prior knowledge and yet, taking $p(\theta) \propto 1$ gives $p(\beta) \propto 1/\beta^2$, while $p(\theta) \propto 1/\theta$ correctly leads to $ p(\beta) \propto 1/\beta$ \cite{prosper1993}.

To map a measurement outcome $E$ into a temperature, we build a point estimator $\tilde{\theta}(E)$. Its quality is assessed via some deviation function $\mathcal{D}[\tilde{\theta}(E), \theta]$ gauging the deviation of $\tilde{\theta}(E)$ from $\theta$. Since all the required information is contained in the joint probability $p(E, \theta) = p(\theta) p(E|\theta) \propto p(E\vert\theta)/\theta $ we write the average uncertainty of $\tilde{\theta}(E)$ as
\begin{equation} 
\bar{\epsilon}_{\mathcal{D}} \coloneqq \int dE\,d\theta \hspace{0.2em} p(E, \theta) \hspace{0.2em} \mathcal{D}[\tilde{\theta}(E), \theta].
\label{generalErr}
\end{equation}
The integration over $\theta$ accounts for all the available prior information and makes Eq.~\eqref{generalErr} temperature-independent. This is a key feature that, unlike the local approach, leads to well-posed optimisation problems \cite{jaynes2003}.

Our next step is to establish the form of the deviation function $\mathcal{D}$. Let the dimensionless scalar $x \in (-\infty, \infty) $, and let its prior be $p(x)\propto 1$. As discussed, e.g., in \cite{jaynes2003}, this is a translationally invariant estimation problem, for which a flat prior represents complete ignorance. The deviation of $\tilde{x}(E)$ from $x$ is then naturally quantified by the $k$-distance $\mathcal{D}[\tilde{x}(E),x]=|\tilde{x}(E)-x|^k$. Now we observe that setting $ x = \alpha \log(k_B \theta/\varepsilon_0) $ maps this hypothetical scenario into our thermometry problem, since $ p(x)dx = p(\theta)d\theta$ implies $ p(x) \propto 1 \mapsto p (\theta) \propto 1/\theta $. Here, $ \varepsilon_0 $ is an arbitrary constant with units of energy included merely for dimension neutralisation \cite{matta2011}, while $\alpha$ is a proportionality factor. Therefore, 
\begin{equation}
\mathcal{D}[\tilde{x}(E),x] \mapsto \mathcal{D}[\tilde{\theta}(E), \theta] = \Bigg\vert \alpha \log{\left[\frac{\tilde{\theta}(E)}{\theta}\right]}\Bigg\vert^k,
\label{logDevGeneral}
\end{equation}
which is a \textit{bona fide} scale parameter deviation function: it is symmetric, i.e., $\mathcal{D}(\tilde{\theta}, \theta)=\mathcal{D}(\theta, \tilde{\theta})$; it respects the invariance of the problem, that is, $\mathcal{D}(\gamma \tilde{\theta}, \gamma \theta)=\mathcal{D}(\tilde{\theta}, \theta)$; it reaches its absolute minimum at $\tilde{\theta} = \theta$ where it vanishes; and it grows (decreases) monotonically from (towards) that minimum when $\tilde{\theta} > \theta$ ($\tilde{\theta} < \theta$). While there may be other functions compatible with these conditions, Eq.~\eqref{logDevGeneral} is certainly a suitable choice for thermometry. Below we further show that for $\alpha = 1$ and $k=2$, the global framework can be reduced to local thermometry assuming one does have prior (local) information. For that reason, we will fix $\alpha = 1$ and $k=2$ in the following, and after Eq.~\eqref{logDevGeneral} is inserted in Eq.~\eqref{generalErr}, we arrive at  
\begin{equation} 
\bar{\epsilon}_{\mathrm{mle}} = \int dE\,d\theta \hspace{0.2em} p(E, \theta)\,\log^2\left[\frac{\tilde{\theta}(E)}{\theta}\right].
\label{mleErr}
\end{equation}
We call $\bar{\epsilon}_{\mathrm{mle}}$ \emph{mean logarithmic error}. 

\vspace{0.5em}
\noindent\emph{\textbf{Optimal global strategy.---}} Our goal is to find the temperature estimator that is optimal with respect to Eq.~\eqref{mleErr}. To do this, we must minimise $\bar{\epsilon}_{\mathrm{mle}}$ over all possible estimators. Since $\bar{\epsilon}_{\mathrm{mle}}$ is a functional of $ \tilde{\theta}(E) $, i.e., $\bar{\epsilon}_{\mathrm{mle}} = \epsilon[\tilde{\theta}(E)]$, this is achieved by solving the variational problem
\begin{equation}
\delta \epsilon[\tilde{\theta}(E)] = \delta \int dE\,\mathcal{L}[\tilde{\theta}(E), E] = 0,
\label{variationalProb}
\end{equation}
where $\mathcal{L}[\tilde{\theta}(E), E] \coloneqq \int d\theta \hspace{0.1em} p(E, \theta) \hspace{0.1em} \log^2{[\tilde{\theta}(E)/\theta]}$ plays the role of a Lagrangian. We find that the optimal estimator $\tilde{\vartheta}(E)$ minimising Eq.~\eqref{mleErr} is given by 
\begin{equation}
\frac{k_B \tilde{\vartheta}(E)}{\varepsilon_0} = \exp\left[ \int d\theta\, p(\theta|E) \log{\left(\frac{k_B \theta}{\varepsilon_0}\right)} \right], 
\label{optEst}
\end{equation}
where $ p(\theta|E) = p(E,\theta)/p(E) $ is the posterior probability, given by Bayes theorem, and $p(E) = \int d\theta\,p(E,\theta) $ \footnote{Note that Eq.~\eqref{optEst} is independent of the specific value of $k_B/\varepsilon_0$}. Inserting the optimal estimator $\tilde{\vartheta}(E)$ as $\tilde{\theta}(E)$ in Eq.~\eqref{mleErr} further gives the optimal logarithmic error $\bar{\epsilon}_{\mathrm{opt}}$. The latter can be interpreted intuitively when split as
\begin{align}
\bar{\epsilon}_{\mathrm{opt}} = \bar{\epsilon}_\mathrm{p} - \mathcal{K}, 
\label{optUncertainty}
\end{align}
where $\bar{\epsilon}_\mathrm{p} \coloneqq \int d\theta \hspace{0.1em}  p(\theta) \hspace{0.1em} \log^2{(\tilde{\vartheta}_{\mathrm{p}}/\theta)}$ is the uncertainty prior to any measurement, with optimal (prior) estimate $k_B \tilde{\vartheta}_{\mathrm{p}}/ \varepsilon_0 =  \exp[ \int d\theta \hspace{0.1em} p(\theta) \hspace{0.1em} \log{(k_B \theta/\varepsilon_0)}]$, and \begin{equation}
\mathcal{K} \coloneqq \int dE\,p(E)\hspace{0.1em} \log^2{\left[\frac{\tilde{\vartheta}(E)}{\tilde{\vartheta}_\mathrm{p}}\right]}
\end{equation}
can be thought of as the maximal information provided by the measurement $E$, on average. The calculations leading to Eqs.~\eqref{optEst} and \eqref{optUncertainty} are given in the Supplemental Material.  

Eqs.~\eqref{optEst} and \eqref{optUncertainty} constitute our main result. The former gives an optimal estimator, $\tilde{\vartheta}(E)$, requiring no prior assumptions and directly applicable on a given dataset. Eq.~\eqref{optUncertainty} indicates the corresponding uncertainty, $\bar{\epsilon}_{\mathrm{opt}}$. Since $\bar{\epsilon}_{\mathrm{opt}}$ is a true minimum, Eq.~\eqref{optUncertainty} also serves as a generalised precision `bound' for temperature estimation \footnote{An analogous result exists for multi-phase estimation \cite{jesus2020mar}.}. Unlike the CRB \cite{mehboudi2019}, this \emph{holds for any estimator, biased or unbiased}. In addition, these results, currently written for a single measurement with outcome $E$, can be trivially adapted to account for any number of repetitions, i.e., i.i.d. $(E_1, E_2, \dots)$ \cite{jesus2020mar}. \emph{Global thermometry is thus able to build temperature estimates drawn from abitrary datasets, including cases with scarce data}. 

\vspace{0.5em}
\noindent\emph{\textbf{Recovery of local thermometry.---}} While, for the sake of generality, $T$ was initially assumed to be completely unknown, one may insert a more localised prior $p(\theta)$ from Eq.~\eqref{mleErr} onwards \cite{demkowicz2011, jesus2017}. In that case, the hypothesis $\theta$ will effectively lie in a narrow range and the estimator $ \tilde{\theta}(E)$ will be close to $\theta$. One then has $\log^2{[\tilde{\theta}(E)/\theta]} \simeq [\tilde{\theta}(E)/\theta-1]^2$, so that Eq.~\eqref{mleErr} can be approximated as
\begin{equation} 
\bar{\epsilon}_{\mathrm{mle}} \simeq \int d\theta \hspace{0.2em} p(\theta) \hspace{0.2em} \frac{\Delta \tilde{\theta}^2}{\theta^2},
\label{mleApproxErr}
\end{equation}
where $\Delta \tilde{\theta}^2 \coloneqq \int dE\,p(E|\theta) \hspace{0.1em} [\tilde{\theta}(E) - \theta]^2$. Eq. (\ref{mleApproxErr}) is the averaged noise-to-signal ratio $\Delta \tilde{\theta}^2/\theta^2$ weighted by the prior \footnote{Interestingly, $\log^2{[\tilde{\theta}(E)/\theta]}$ and $[\tilde{\theta}(E)-\theta]^2/\theta^2$, which give rise to Eqs.~\eqref{mleErr} and \eqref{mleApproxErr}, respectively, have been previously considered as independent choices with different properties \cite{norstrom1996}. Instead, here we derive Eq.~\eqref{mleErr} by imposing physically motivated constraints, while Eq.~\eqref{mleApproxErr} appears as a local limiting case. Hence, we may regard the former as more fundamental.}. Since $\Delta \tilde{\theta}^2$ is the `frequentist' mean square error \cite{rafal2015}, it may be lower bounded as \cite{kay1993, jaynes2003}     
\begin{equation}
\Delta \tilde{\theta}^2 \geqslant \frac{1}{F(\theta)} \left[1+\frac{\partial b(\theta)}{\partial \theta}\right]^2 + b(\theta)^2,
\label{CRB}
\end{equation}        
where we have introduced the Fisher information
\begin{equation}
F(\theta) = \int \frac{dE}{p(E|\theta)} \left[\frac{\partial p(E|\theta)}{\partial \theta} \right]^2,
\label{FisherInfo}
\end{equation}
and the bias $b(\theta) \coloneqq \int dE\,p(E|\theta) \hspace{0.1em} [\tilde{\theta}(E) -\theta]$. Provided that $b(\theta) \simeq 0$ as is assumed in the local approach \cite{rafal2015}, we see that Eqs.~\eqref{mleApproxErr}, \eqref{CRB}, and $p(\theta)/\theta^2 \geqslant 0$ lead to the Cram\'{e}r--Rao-like bound
\begin{equation}
\bar{\epsilon}_{\mathrm{mle}} \gtrsim \int d\theta\,\frac{p(\theta)}{\theta^2 F(\theta)} \coloneqq \bar{\epsilon}_\mathrm{cr}.
\label{asympApprox}
\end{equation} 
Equality would hold only inasmuch as the assumptions underpinning Eq.~\eqref{asympApprox} are fulfilled. Namely, $b(\theta)\simeq 0$ and closeness of $\tilde{\theta}(E)$ and $\theta$, which makes $\bar{\epsilon}_\mathrm{cr}$ a local quantifier even when $\theta$ is integrated. Hence, we have derived a local form of thermometry as a limit of the global framework.

Note that the quantifier $ \bar{\epsilon} = \int d\theta\, p(\theta) F(\theta)^{-1} $ has been proposed as an attempt to supersede the local paradigm \cite{pearce2017, mok2020}. Notwithstanding its merits---Ref.~\cite{mok2020} reports results beyond standard local thermometry---such $ \bar{\epsilon} $ is \emph{not} scale invariant. This might be ignored if the prior is narrowly concentrated around some fixed $\theta'$, since then $ \bar{\epsilon}_{\mathrm{cr}} \simeq [(\theta')^2 F(\theta')]^{-1}$ and  $\bar{\epsilon} \simeq F(\theta')^{-1} $. That is, one is left in both cases with the Fisher information. But this limiting assumption is unnecessary in our truly global and intrinsically scale invariant approach. 

\begin{figure}[t]
\centering
\includegraphics[trim={0.1cm 0.1cm 1.5cm 0.6cm},clip,width=\linewidth]{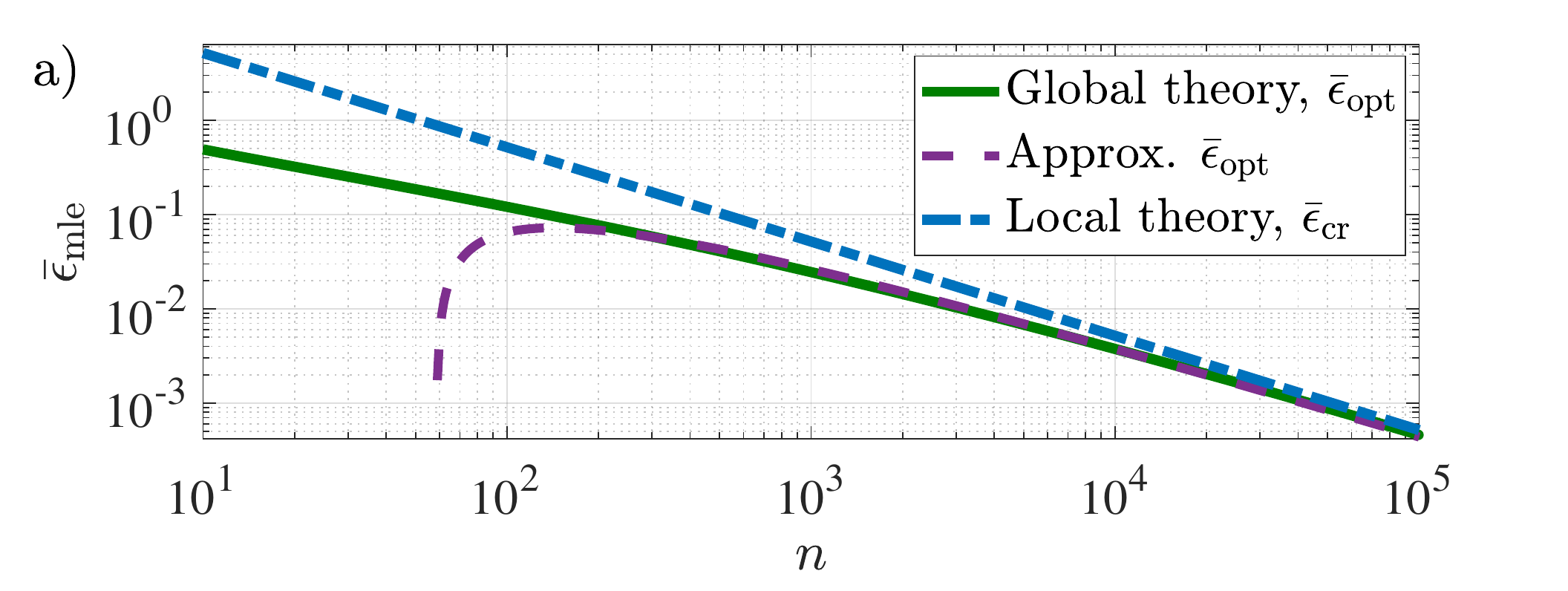}
\includegraphics[trim={0.1cm 0.1cm 1.5cm 0.7cm},clip,width=\linewidth]{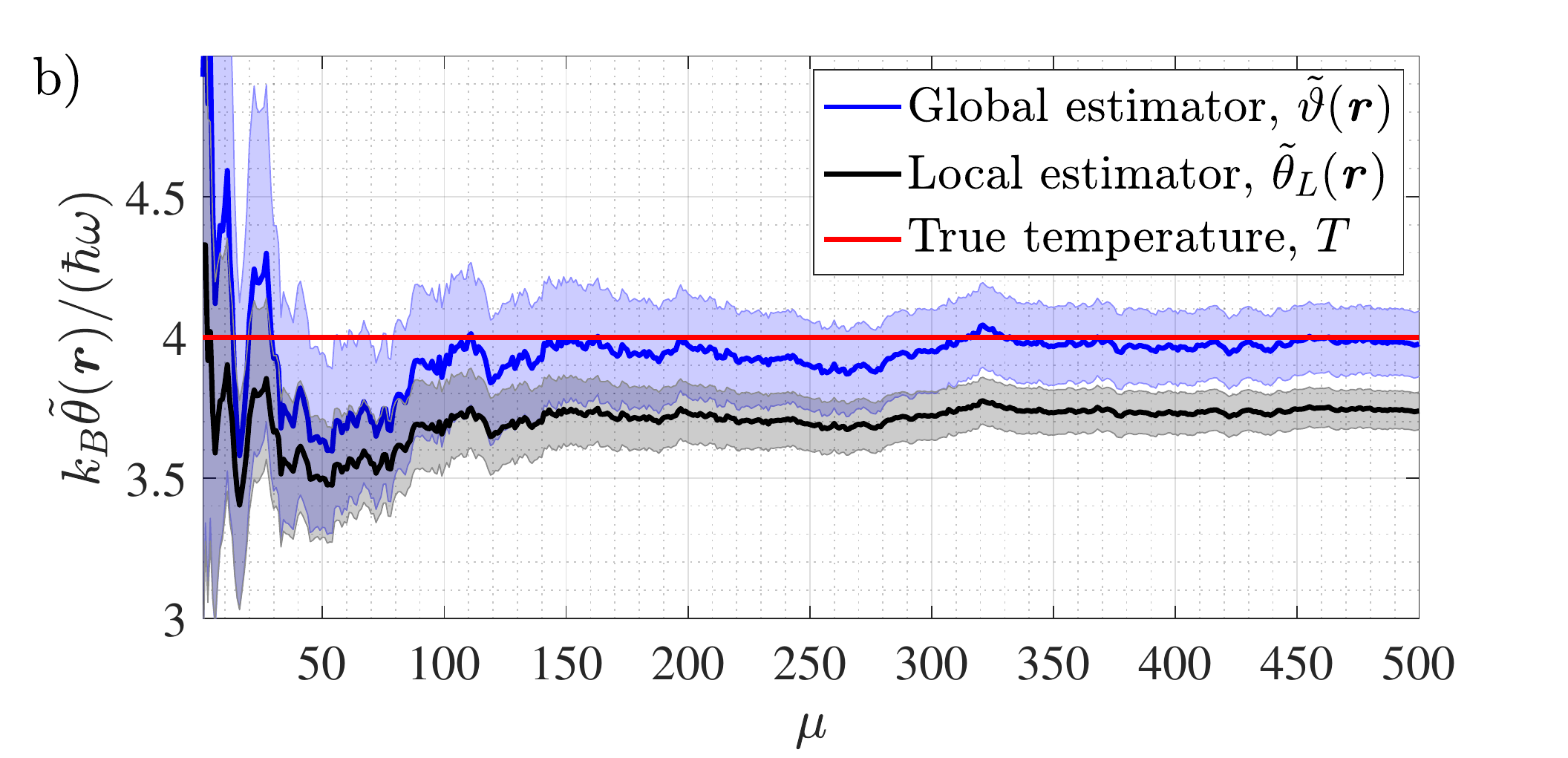}
	\caption{(colour online) a) Log-log plot of the global optimum $\bar{\epsilon}_{\mathrm{opt}}$ in Eq.~\eqref{optUncertainty} (solid green) and local Cram\'{e}r--Rao-like bound in Eq.~\eqref{asympApprox} (dot-dashed blue) for a gas of $ n $ non-interacting spin-$1/2$ particles in thermal equilibrium. As can be seen, the global optimum is lower than the local bound unless $n \rightarrow \infty$, meaning that the local bound misses information when $n$ is `too small'. Eq.~(\ref{secondTermBayes}) gives an asymptotic expansion matching the global optimum when $n \gtrsim 10^2$ (dashed purple). b) Data analysis in global thermometry. We simulated the outcomes $\boldsymbol{r} = (r_1, \dots, r_\mu)$ of $\mu$ energy measurements on the $n$-spin gas, with $n = 150$ and true temperature (solid red) $k_B T /(\hbar\omega) = 4$. We then post-processed $\boldsymbol{r}$ using the global estimator in Eq.~\eqref{optEst} (dotted blue) and a local estimator initialised at $k_B \theta_0 /(\hbar\omega) = 3$ \cite{kay1993, rafal2015} (dot-dashed black). The global estimate converges to the true temperature after $\mu \simeq 10^2$ shots. In contrast, the local theory leads to a biased estimate even for $\mu \simeq 500$. The global error is calculated via Eq.~\eqref{mleErr} but with average over the posterior $p(\theta|\boldsymbol{r})$ (see Supplemental Material). The standard CRB gives the local error \cite{kay1993, rafal2015}.} 
\label{spinBayes}
\end{figure}

\vspace{0.5em}
\noindent\emph{\textbf{Example: Non-interacting spin gas.---}} We now turn to illustrate how to find the best thermometric precision one could possibly get. Let us consider a gas of $n$ non-interacting spin-1/2 particles with energy gap $\hbar \omega$ in thermodynamic equilibrium. This could correspond, e.g., to a dilute cloud of impurities fully equilibrated with a co-trapped ultra-cold majority gas, whose temperature needs to be measured precisely \cite{bouton2019single,olf2015thermometry,mehboudi2019using,mitchison2020}. The probability of measuring the total energy $E$ to be $ r \hbar\omega $, with $r=0, 1, \dots, n$, is \cite{jahnke2011, wyllie1970}
\begin{equation}
p(r|\theta) = \binom{n}{r} \frac{\exp[- r \hbar \omega/(k_B \theta)]}{Z[\hbar \omega /(k_B\theta)]},
\label{likelihoodFermions}
\end{equation}
where $Z[\hbar \omega /(k_B\theta)] = \lbrace \exp[-\hbar \omega/(k_B \theta)]+1\rbrace^{n}$ is the partition function. Additionally, suppose that $p(\theta) \propto 1/\theta$ is defined, for instance, on the finite support $k_B \theta / (\hbar \omega) \in [0.1, 10]$, so that normalisation gives $p(\theta) = 1/[2 \theta \log(10)]$. 

The optimum $\bar{\epsilon}_{\mathrm{opt}}$ can be readily evaluated after inserting the prior $p(\theta)$ and the likelihood \eqref{likelihoodFermions} into Eq.~\eqref{optUncertainty} \footnote{Numerical algorithms on GitHub \href{https://github.com/jesus-rubiojimenez/QuThermometry-global}{https://github.com/jesus-rubiojimenez/QuThermometry-global} (2020)}. The result, for $ n $ ranging from $ 10 $ to $ 10^5 $, is shown in Fig.~\ref{spinBayes}~a). The local limit of this error [cf. Fig.~\ref{spinBayes}~a)] is shown to take the form $\bar{\epsilon}_{\mathrm{cr}} \simeq 51.7/n$ in the Supplemental Material. Comparing both, we observe convergence as $n \rightarrow \infty$, confirming the emergence of local thermometry within the global theory. However, for finite $n$ we see that $\bar{\epsilon}_{\mathrm{cr}} > \bar{\epsilon}_\mathrm{opt} $. That is, while the global estimator extracts all available information for any $n$, a local approach leads to information loss when $n$ is small.

To gain analytical insight into the scaling of $\bar{\epsilon}_{\mathrm{opt}}$ with $n$, first we note that, in this specific example, $\bar{\epsilon}_\mathrm{cr}-\bar{\epsilon}_\mathrm{opt} \simeq b \hspace{0.1em} n^q$. A numerical fit \footnote{Specifically, a Bayesian fit using $\log[\bar{\epsilon}_{\mathrm{cr}}(n)-\bar{\epsilon}_{\mathrm{opt}}(n)] = q \log(n) + \log(b) + e(n)$, with Gaussian errors $e(n)$.} renders the asymptotic expansion 
\begin{equation}
\bar{\epsilon}_\mathrm{opt} \simeq \frac{51.7}{n} - \frac{143}{n^{5/4}}
\label{secondTermBayes},
\end{equation}
which, as shown in Fig.~\ref{spinBayes}~a), matches the true optimum $\bar{\epsilon}_{\mathrm{opt}}$ when $n \gtrsim 10^2$. This in turn allows to assess how large $ n $ needs to be for the (local) $\sim 1/n$ scaling to hold approximately. Consider the equation $|\bar{\epsilon}_\mathrm{cr}[n(\tau)] - \bar{\epsilon}_\mathrm{opt}[n(\tau)]|= \tau \hspace{0.2em} \bar{\epsilon}_\mathrm{opt}[n(\tau)]$ \cite{jesus2017}, where $n(\tau)$ is the minimum number of spins which drives the relative error between the optimal and asymptotic curves below a tolerance $\tau$. Using Eq.~\eqref{secondTermBayes} gives the condition $n(\tau) \simeq 58.5\,(1+1/\tau )^4$. As expected, $n(\tau) \rightarrow \infty$ when $\tau \rightarrow 0$. But, even when tolerating a $5\%$ deviation, spin numbers of $n \sim 10^7 $ are required for local thermometry to give a correct scaling. Admittedly, the value of $n(\tau)$ is protocol-dependent \cite{jesus2017, jesus2018}, but even a simple example suffices to illustrate the perils of the local framework---information loss (Fig.~\ref{spinBayes}~a)) and failure to produce a valid low-$n$ scaling (Eq.~\eqref{secondTermBayes}).  

\vspace{0.5em}
\noindent\emph{\textbf{Data analysis in global thermometry.---}} The global framework does not only enable a more comprehensive picture of fundamental limits, but, perhaps more importantly, is also a reliable tool for experimental data analysis. Consider the following protocol: a gas of $n$ spin-1/2 particles is prepared; its energy $r\hbar\omega$ is measured; both steps are repeated $\mu$ times, generating the outcomes $(r_1, r_2, \dots, r_{\mu}) \coloneqq \boldsymbol{r}$. The global estimate $\tilde{\vartheta}(\boldsymbol{r})$ in Eq.~\eqref{optEst} is calculated using the posterior $ p(\theta|\boldsymbol{r}) \propto p(\boldsymbol{r}|\theta)/\theta$, with $k_B\theta/(\hbar \omega) \in [0.1, 10]$, $p(\boldsymbol{r}|\theta) = \prod_{i=1}^\mu p(r_i|\theta)$ and $p(r_i|\theta)$ given by Eq.~\eqref{likelihoodFermions}. To assess its uncertainty, the average over $p(\boldsymbol{r}, \theta)$ in Eq.~\eqref{mleErr} is instead taken over $p(\theta|\boldsymbol{r})$, since, in experiments, the outcomes $\boldsymbol{r}$ are known \cite{jaynes2003, jesus2019thesis}. The resulting error $\bar{\epsilon}_{\mathrm{mle}}(\boldsymbol{r})$ is obviously outcome-dependent, while temperature-independent \footnote{In the Supplemental material we prove that Eq.~\eqref{optEst} is optimal also with respect to $\bar{\epsilon}_{\mathrm{mle}}(\boldsymbol{r})$.}. Recalling that the logarithmic error is a noise-to-signal ratio, we introduce the Bayesian `error bar' $\Delta \tilde{\theta}(\boldsymbol{r}) \coloneqq \tilde{\theta}(\boldsymbol{r}) \sqrt{\bar{\epsilon}_{\mathrm{mle}}(\boldsymbol{r})}$. This analysis may be compared with the local CRB-based estimator $\tilde{\theta}(\boldsymbol{r})_L = \theta_0 + [\mu F(\theta_0)]^{-1}\,\partial \log[p(\boldsymbol{r}|\theta_0)]/\partial \theta$, with error $\Delta \tilde{\theta}_L = 1/\sqrt{\mu F(\theta_0)}$ \cite{kay1993, rafal2015}. Here, $\theta_0$ is an initial `hint' at the true temperature $T$, a prerequisite in local thermometry.

We simulated the outcomes $\boldsymbol{r}$ arising from the aforementioned protocol, for a gas with $n = 150$ spins and true temperature $k_B T/(\hbar \omega) = 4$. The local estimate $\tilde{\theta}_{L} \pm \Delta\tilde{\theta}_{L}$, initialised with the `hint' $k_B \theta_0/(\hbar \omega) = 3$, is shown in Fig.~\ref{spinBayes}~b) (dot-dashed black) to be biased even when $\mu \gg 1$. This contrasts with the convergence of the global estimate $\tilde{\vartheta}\pm \Delta \tilde{\vartheta}$ (dotted blue) to the true temperature (solid red). One could argue for a `two-step' method where a part of the data is used to provide a better $\theta_0$ prior to applying local thermometry. Yet, one cannot anticipate how many trials a good seed $\theta_0$ requires, nor when such $\theta_0$ is sufficiently close to the true temperature. The global framework is instead general, reliable and works at once. Further evidence is provided in the Supplemental Material, where a comparison between global thermometry and a histogram-fitting procedure demonstrates the potential gain of global techniques for experiments with limited data.

\vspace{0.5em}
\noindent\emph{\textbf{Conclusions.---}} We have demonstrated that local precision benchmarks are insufficient whenever few data or no substantial prior knowledge are available. On the contrary, \emph{a global approach is applicable to any temperature-estimation protocol regardless of the measurement record length, and can naturally account for any degree of prior information}. For instance, it would be interesting to exploit Eq.~\eqref{optEst} to post-process data measured in the nanokelvin regime, which is experimentally accessible with ultracold Bose and Fermi gases \cite{javanainen1995,leanhardt2003cooling,ruostekoski2009,olf2015thermometry,2016onofrio,bouton2019single,carcy2021} and relevant for quantum simulation \cite{bloch2012quantum}. In addition, since Eq.~\eqref{mleErr} is deduced at the level of probability distributions (i.e., with no explicit consideration of the Born rule $p(E|\theta) = \mathrm{Tr}[\Pi(E)\rho(\theta)$] \footnote{Here $\Pi$ denotes a probability-operator measurement (POM), while $\rho$ is a state operator.}), Eq.~\eqref{optEst} can be applied also to classical systems. Finally, note that the key assumption behind Eq.~\eqref{mleErr} is that the parameter is a scale, which makes it applicable beyond temperature estimation (e.g., to estimate biochemical rates in single-molecule experiments \cite{baaske2014, subramanian2020, subramanian2021, eerqing2021}). 

From a fundamental perspective, combining Eq.~\eqref{optEst}---the rule to calculate optimal estimates---with the theoretical optimum \eqref{optUncertainty} provides a powerful tool to address open problems in thermometry. These include pushing precision limits further by optimising over the energy spectrum of the probe system \cite{correa2015, mok2020}, or over measured quantities beyond energy (see Supplemental Material). Moreover, the global formalism may be extended to accommodate for the non-equilibrium states \cite{de2016local,hovhannisyan2018,potts2019} resulting from limited access to interacting thermalised probes, and gives the theoretical support needed to derive more fundamental energy--temperature uncertainty relations \cite{miller2018}. Global quantum thermometry has thus potential to become the new standard for thermometry in the quantum regime.

%--------------------------------------

\begin{acknowledgments}

\vspace{0.5em}
\noindent{\textit{\textbf{Acknowledgements.---}}} We thank A. Bayat, W.-K. Mok, K. Bharti, M. Perarnau-Llobet, A. Luis and S. Subramanian for helpful discussions. J.R. and J.A. acknowledge support from EPSRC (Grant No. EP/T002875/1), and J.A. acknowledges support from EPSRC (Grant No. EP/R045577/1) and the Royal Society.

\end{acknowledgments}

% References
\bibliography{refs2021jun}

\onecolumngrid
\setcounter{equation}{0}
\renewcommand\theequation{S\arabic{equation}}

\section{Supplemental material}

\subsection{Optimisation of the mean logarithmic error}

In this work we employ the mean logarithmic error
\begin{equation}
\bar{\epsilon}_{\mathrm{mle}} =
  \int dE\,d\theta\,p(E, \theta) \hspace{0.1em} \mathrm{log}^2\left[\frac{\tilde{\theta}(E)}{\theta}\right] \coloneqq \epsilon[\tilde{\theta}(E)] 
\label{mleVarProbSupp}
\end{equation}
as the figure of merit. To optimise it with respect to all possible estimators, first we solve the variational problem 
\begin{equation}
\delta \epsilon[\tilde{\theta}(E)] = 0,
\label{variationalProbSupp}
\end{equation}
which is mathematically equivalent to require that \cite{mathematics2004}
\begin{equation}
\frac{d}{d\alpha} \epsilon[\tilde{\theta}(E)+\alpha \tilde{\eta}(E)] \bigg\rvert_{\alpha = 0} = 0,~~\text{for~all}~~\tilde{\eta}(E). 
\label{mathVarProblem}
\end{equation}
By inserting Eq. (\ref{mleVarProbSupp}) in the left hand side of Eq. (\ref{mathVarProblem}), and using the fact that $p(E, \theta) = p(E) p(\theta|E)$, we find that
\begin{align}
\frac{d}{d\alpha} \epsilon[\tilde{\theta}(E)+\alpha \tilde{\eta}(E)]  \bigg\rvert_{\alpha = 0} & =
2  \int dE\,d\theta\,p(E, \theta)\,\mathrm{log}\left[\frac{\tilde{\theta}(E)+\alpha \hspace{0.1em} \tilde{\eta}(E)}{\theta}\right] \frac{\tilde{\eta}(E)}{\tilde{\theta}(E) + \alpha\,\tilde{\eta}(E)} \bigg\rvert_{\alpha = 0}
\nonumber \\
& =  2  \int dE\,\frac{p(E)}{\tilde{\theta}(E)}\int d\theta\,p(\theta|E)\,\mathrm{log}\left[\frac{\tilde{\theta}(E)}{\theta}\right] \tilde{\eta}(E),
\end{align}
which vanishes for all variations $\tilde{\eta}(r)$ when 
\begin{equation}
\int d\theta\,p(\theta|E)\,\mathrm{log}\left[\frac{\tilde{\theta}(E)}{\theta}\right] = 0. 
\label{varCond}
\end{equation}
This condition is an equation for $\tilde{\theta}(E)$, and its solution can be found straightforwardly if we first rewrite the logarithmic factor as
\begin{equation}
\mathrm{log}\left[\frac{\tilde{\theta}(E)}{\theta}\right] = \mathrm{log}\left[\frac{k_B \tilde{\theta}(E)}{\varepsilon_0} \right]-\mathrm{log}\left( \frac{k_B \theta}{\varepsilon_0}\right),
\end{equation}
arriving at
\begin{equation}
\frac{k_B \tilde{\theta}(E)}{\varepsilon_0} 
= \exp\left[\int d\theta\,p(\theta|E)\log\left(\frac{k_B \theta}{\varepsilon_0}\right) \right] \coloneqq \frac{k_B \tilde{\vartheta}(E)}{\varepsilon_0}.
\label{optEstSupp}
\end{equation}
In other words, we have proven that $\tilde{\vartheta}(E)$ makes the mean logarithmic error extremal.

Next we wish to verify that $\tilde{\theta}(E) = \tilde{\vartheta}(E)$ is a minimum, which we may do via the functional version of the second derivative test. Upon calculating the second variation of the error, evaluated at $\tilde{\theta}(E) = \tilde{\vartheta}(E)$, we find that
\begin{align}
\frac{d^2}{d\alpha^2} \epsilon[\tilde{\vartheta}(E)+\alpha \tilde{\eta}(E)] \bigg\rvert_{\alpha = 0}  & = 
2  \int dE\,d\theta\,p(E, \theta)\,\left\lbrace 1 - \mathrm{log}\left[\frac{\tilde{\vartheta}(E)+\alpha \hspace{0.1em} \tilde{\eta}(E)}{\theta}\right]\right\rbrace \left[\frac{\tilde{\eta}(E)}{\tilde{\vartheta}(E) + \alpha \hspace{0.1em} \tilde{\eta}(E)}\right]^2 \bigg\rvert_{\alpha = 0}
\nonumber \\
& = 2  \int dE\,p(E)\,\left[\frac{\tilde{\eta}(E)}{\tilde{\vartheta}(E)}\right]^2 \int d\theta \hspace{0.1em} p(\theta|E) \hspace{0.1em} \left\lbrace 1 - \mathrm{log}\left[\frac{k_B \tilde{\vartheta}(E)}{\varepsilon_0}\right]+\mathrm{log}\left[\frac{k_B \theta}{\varepsilon_0}\right]\right\rbrace
\nonumber \\
& = 2  \int dE\,p(E)\,\left[\frac{\tilde{\eta}(E)}{\tilde{\vartheta}(E)}\right]^2 \left\lbrace 1 - \mathrm{log}\left[\frac{k_B \tilde{\vartheta}(E)}{\varepsilon_0}\right]+\int d\theta\,p(\theta|E)\,\mathrm{log}\left(\frac{k_B \theta}{\varepsilon_0}\right)\right\rbrace
\nonumber \\
& = 2  \int dE\,p(E)\,\left[\frac{\tilde{\eta}(E)}{\tilde{\vartheta}(E)}\right]^2 > 0
\label{secondVar}
\end{align} 
for non-trivial variations (i.e., $\tilde{\eta}(E)\neq 0$); consequently, we conclude that $\tilde{\vartheta}(E)$ is the optimal estimator that minimises the Bayesian uncertainty in Eq. (\ref{mleVarProbSupp}).
 
The minimum mean logarithmic error can now be found by introducing the optimal estimator $\tilde{\vartheta}(E)$ in Eq. (\ref{mleVarProbSupp}). This operation leads to
\begin{align}
 \epsilon[\tilde{\vartheta}(E)] & =  \int dE\,d\theta\,p(E, \theta) \hspace{0.1em} \mathrm{log}^2\left[\frac{\tilde{\vartheta}(E)}{\theta}\right] 
\nonumber \\
& =  \int dE\,p(E)\,\int d\theta\,p(\theta|E)\,\left\lbrace \mathrm{log}\left[\frac{k_B \tilde{\vartheta}(E)}{\varepsilon_0}\right] - \mathrm{log}\left(\frac{k_B \theta}{\varepsilon_0}\right) \right\rbrace^2
\nonumber \\
& =  \int d\theta\,p(\theta)\,\mathrm{log}^2\left(\frac{k_B \theta}{\varepsilon_0}\right) + \int dE\,p(E)\,\mathrm{log}^2\left[\frac{k_B \tilde{\vartheta}(E)}{\varepsilon_0}\right] - 2 \int dE\,p(E)\,\mathrm{log}\left[\frac{k_B \tilde{\vartheta}(E)}{\varepsilon_0}\right] \int d\theta\,p(\theta|E)\,\mathrm{log}\left(\frac{k_B \theta}{\varepsilon_0}\right)
\nonumber \\
& =  \int d\theta\,p(\theta)\,\mathrm{log}^2\left(\frac{k_B \theta}{\varepsilon_0}\right) - \int dE\,p(E)\,\mathrm{log}^2\left[\frac{k_B \tilde{\vartheta}(E)}{\varepsilon_0}\right] 
\coloneqq \bar{\epsilon}_{\mathrm{opt}}.
\label{mmleSupp}
\end{align}

\subsection{Interpretation of the logarithmic optimum}

It is possible to associate a clear meaning to the optimum in Eq. (\ref{mmleSupp}). To achieve this, let us first address a related but different question: given the prior probability $p(\theta)$, and no experimental data, what is the best prior estimate $\tilde{\theta}_\mathrm{p}$ for the true temperature $T$? Since we are assuming that no measurement has been yet performed, the logarithmic uncertainty in Eq.(\ref{mleVarProbSupp}) needs to be substituted by $\int d\theta \hspace{0.1em} p(\theta) \hspace{0.1em} \mathrm{log}^2(\tilde{\theta}_\mathrm{p}/\theta)$, so that $\tilde{\theta}_\mathrm{p}$ is now a number rather than a function. Then, solving the simple optimisation problem
\begin{equation}
\frac{d}{d\tilde{\theta}_\mathrm{p}} \int d\theta\,p(\theta)\,\mathrm{log}^2\left(\frac{\tilde{\theta}_\mathrm{p}}{\theta}\right) = 0, \hspace{1em} \frac{d^2}{d\tilde{\theta}_\mathrm{p}^2} \int d\theta\,p(\theta)\,\mathrm{log}^2\left(\frac{\tilde{\theta}_\mathrm{p}}{\theta}\right) > 0  
\end{equation}
shows that the optimal prior estimate is 
\begin{equation}
\frac{k_B \tilde{\theta}_\mathrm{p}}{\varepsilon_0} = \exp\left[ \int \hspace{0.1em} d\theta\,p(\theta)\,\log \left(\frac{k_B \theta}{\varepsilon_0}\right)\right] \coloneqq \frac{k_B \tilde{\vartheta}_\mathrm{p}}{\varepsilon_0},
\label{priorOptEstSupp}
\end{equation}
with optimal prior uncertainty
\begin{equation}
\bar{\epsilon}_{\mathrm{p}} = \int d\theta\,p(\theta)\,\mathrm{log}^2\left(\frac{\tilde{\vartheta}_\mathrm{p}}{\theta}\right) = \int d\theta\,p(\theta)\,\mathrm{log}^2\left(\frac{k_B \theta}{\varepsilon_0}\right) - \left[\int d\theta\,p(\theta)\,\mathrm{log}\left(\frac{k_B \theta}{\varepsilon_0}\right)\right]^2.
\label{priormmleSupp}
\end{equation}
Note that we have used
\begin{equation}
\mathrm{log}\left(\frac{\tilde{\vartheta}_\mathrm{p}}{\theta}\right) = \mathrm{log}\left(\frac{k_B \tilde{\vartheta}_\mathrm{p}}{\varepsilon_0} \right)-\mathrm{log}\left(\frac{k_B \theta}{\varepsilon_0}\right).
\end{equation}

Returning to our original problem, we can now manipulate the optimum in Eq. (\ref{mmleSupp}) as
\begin{align}
 \bar{\epsilon}_{\mathrm{opt}} = & \int d\theta\,p(\theta)\,\mathrm{log}^2\left(\frac{k_B \theta}{\varepsilon_0}\right)
 - 2\int d\theta\,p(\theta)\,\mathrm{log}\left(\frac{k_B \theta}{\varepsilon_0}\right)\mathrm{log}\left(\frac{k_B \tilde{\vartheta}_p}{\varepsilon_0}\right)
 + \int d\theta\,p(\theta)\,\mathrm{log}^2\left(\frac{k_B \tilde{\vartheta}_p}{\varepsilon_0}\right)
 \nonumber \\
 & - \int dE\,p(E)\,\mathrm{log}^2\left[\frac{k_B \tilde{\vartheta}(E)}{\varepsilon_0}\right] 
 + 2\int d\theta\,p(\theta)\,\mathrm{log}\left(\frac{k_B \theta}{\varepsilon_0}\right)\mathrm{log}\left(\frac{k_B \tilde{\vartheta}_p}{\varepsilon_0}\right)
 - \int d\theta\,p(\theta)\,\mathrm{log}^2\left(\frac{k_B \tilde{\vartheta}_p}{\varepsilon_0}\right)
 \nonumber \\
 = &\int d\theta\,p(\theta)\,\mathrm{log}^2\left(\frac{ \tilde{\vartheta}_p}{\theta}\right) 
 - \int dE\,p(E)\,\mathrm{log}^2\left[\frac{k_B \tilde{\vartheta}(E)}{\varepsilon_0}\right] 
 \nonumber \\
 &  + 2\int dE\,p(E)\,\mathrm{log}\left[\frac{k_B \tilde{\vartheta}(E)}{\varepsilon_0}\right]\mathrm{log}\left(\frac{k_B \tilde{\vartheta}_p}{\varepsilon_0}\right)
 - \int dE\,p(E)\,\mathrm{log}^2\left(\frac{k_B \tilde{\vartheta}_p}{\varepsilon_0}\right)
 \nonumber \\
 = &\int d\theta\,p(\theta)\,\mathrm{log}^2\left(\frac{ \tilde{\vartheta}_p}{\theta}\right)
 - \int dE\,p(E)\,\mathrm{log}^2\left[\frac{\tilde{\vartheta}(E)}{\tilde{\vartheta}_\mathrm{p}}\right] 
 \coloneqq \bar{\epsilon}_{\mathrm{p}} - \mathcal{K},
 \label{optimumInterSupp}
\end{align}
where we have employed Eq.~(\ref{priormmleSupp}), the fact that $\int d\theta\,p(\theta) = \int dE\,p(E) = 1$ and
\begin{equation}
\int d\theta\,p(\theta)\,\mathrm{log}\left(\frac{k_B \theta}{\varepsilon_0}\right)
= \int dE\,p(E)\,\int d\theta \hspace{0.1em}  p(\theta|E) \hspace{0.1em} \mathrm{log}\left(\frac{k_B \theta}{\varepsilon_0}\right) 
= \int dE\,p(E)\,\mathrm{log}\left[\frac{k_B \tilde{\vartheta}(E)}{\varepsilon_0}\right],
\end{equation}
which stems from Eq.~(\ref{optEstSupp}). According to the last line in Eq. (\ref{optimumInterSupp}), the quantity $\mathcal{K}$ is a logarithmic distance between the optimal estimator $\tilde{\vartheta}(E)$ and the optimal prior estimate $\tilde{\vartheta}_\mathrm{p}$. As such, the more different from $\tilde{\vartheta}_\mathrm{p}$ the optimal estimator $\tilde{\vartheta}(E)$ is, the larger their logarithmic distance becomes, which reduces the optimal error $\bar{\epsilon}_{\mathrm{opt}}$ because $\mathcal{K}$ is subtracted from the prior uncertainty $\bar{\epsilon}_{\mathrm{p}}$. Hence, we may interpret $\mathcal{K}$ as a measure of the maximal information that the measurement outcome $E$ can provide on average with respect to the information that was already available prior to performing the experiment. 

\subsection{Bayesian analogue of the noise-to-signal ratio}

In the main text we argue that the notion of locality exploited in local estimation theory \cite{rafal2015} may be implemented in the Bayesian framework by imposing that $\tilde{\theta}(E)$ is generally close to $\theta$. In that case, and using that $p(E,\theta) = p(\theta) p(E|\theta)$, the mean logarithmic error in Eq. (\ref{mleVarProbSupp}) can be Taylor-expanded around $\tilde{\theta}(E) = \theta$ and approximated as
\begin{equation}
\bar{\epsilon}_{\mathrm{mle}} 
\simeq \int dE\,d\theta\,p(E, \theta)\,\left[\frac{\tilde{\theta}(E)}{\theta} - 1\right]^2 
= \int d\theta\,\frac{p(\theta)}{\theta^2} \int dE\,p(E|\theta) \left[\tilde{\theta}(E) - \theta \right]^2 
\coloneqq \int d\theta\,p(\theta)\,\frac{\Delta \tilde{\theta}^2}{\theta^2},
\label{BayesSNR}
\end{equation}
which is a Bayesian analogue of the noise-to-signal ratio $\Delta \tilde{\theta}^2/\theta^2$. This result differs from that by Phillies \cite{phillies1984} and Prosper \cite{prosper1993}, who instead constructed a Bayesian noise-to-signal ratio by using the moments of the posterior probability $p(\theta|E)$. However, this ignores the necessity of acknowledging the nature of the unknown parameter not only via the prior, but also when choosing the deviation function, while our method takes both of these into account. 

\subsection{Asymptotics for a gas of non-interacting spins-$1/2$ particles}

Let 
\begin{equation}
p(r|\theta) = \binom{n}{r} \frac{\exp[- r \hbar \omega/(k_B \theta)]}{Z[\hbar \omega /(k_B\theta)]}
\label{likelihoodFermionsSupp}
\end{equation}
be the probability, conditioned on the hypothesis $\theta$, of measuring the total energy of a gas of $n$ non-interacting spins-$1/2$ particles to be $r\hbar \omega$ \cite{jahnke2011, wyllie1970}, with $r=0, 1, \dots, n$ and partition function
\begin{equation}
Z\left(\frac{\hbar \omega}{k_B\theta}\right) = \sum_{r=0}^n \binom{n}{r} \exp\left(- \frac{r \hbar \omega}{k_B \theta}\right) = \sum_{r=0}^n \binom{n}{r} \left[\exp\left(- \frac{\hbar \omega}{k_B \theta}\right)\right]^r 1^{(n-r)} = \left[\exp\left(- \frac{\hbar \omega}{k_B \theta}\right) + 1 \right]^n.
\end{equation}
According to our results in the main text, the calculation of the Bayesian optimum in Eq. (\ref{mmleSupp}) may be approximated as
\begin{equation}
\bar{\epsilon}_{\mathrm{opt}} \simeq  \bar{\epsilon}_{\mathrm{cr}} = \int d\theta\,\frac{p(\theta)}{\theta^2 F(\theta)}
\label{asymSupp}
\end{equation}
when $n \gg 1$. In our case, the Fisher information becomes
\begin{align}
F(\theta) & = \sum_r \frac{1}{p(r|\theta)}\left[\frac{\partial p(r|\theta)}{\partial \theta} \right]^2
\nonumber \\
& = \sum_{r=0}^n \binom{n}{r} \exp\left(\frac{r \hbar \omega}{k_B \theta}\right) Z\left(\frac{\hbar \omega}{k_B\theta}\right) \left( \frac{\partial}{\partial \theta} \left\lbrace\frac{\exp\left[- r \hbar \omega/(k_B \theta) \right]}{Z[\hbar \omega /(k_B\theta)]} \right\rbrace \right)^2
\nonumber \\
& = \sum_{r=0}^n \binom{n}{r} \frac{\exp\left[- r \hbar \omega/(k_B \theta) \right]}{Z[\hbar \omega /(k_B\theta)]} \left\lbrace\frac{r\hbar \omega}{k_B \theta^2} - \frac{Z'[\hbar \omega /(k_B\theta)]}{Z[\hbar \omega /(k_B\theta)]} \right\rbrace^2 
\nonumber \\
& = \left\lbrace\frac{Z'[\hbar \omega /(k_B\theta)]}{Z[\hbar \omega /(k_B\theta)]}\right\rbrace^2 
- \frac{2\hbar \omega}{k_B \theta^2} \frac{Z'[\hbar \omega /(k_B\theta)]}{Z[\hbar \omega /(k_B\theta)]} \sum_{r=0}^n \binom{n}{r} \frac{\exp\left[- r \hbar \omega/(k_B \theta) \right]}{Z[\hbar \omega /(k_B\theta)]} r 
+ \left(\frac{\hbar \omega}{k_B \theta^2}\right)^2\sum_{r=0}^n \binom{n}{r} \frac{\exp\left[- r \hbar \omega/(k_B \theta) \right]}{Z[\hbar \omega /(k_B\theta)]} r^2
\nonumber \\
& = \frac{\exp[\hbar \omega/(k_B \theta)]}{n} \left\lbrace\frac{Z'[\hbar \omega /(k_B\theta)]}{Z[\hbar \omega /(k_B\theta)]} \right\rbrace^2
\nonumber \\
& = \frac{1}{\theta^2}\left(\frac{\hbar \omega}{2 k_B \theta}\right)^2 \frac{n}{\cosh^2[\hbar \omega/(2 k_B \theta)]},
\label{fisherSupp}
\end{align}
where we have used the identities
\begin{align}
\sum_{r=0}^n \binom{n}{r} \exp\left(- \frac{r \hbar \omega}{k_B \theta}\right) r & = \frac{k_B \theta^2}{\hbar \omega} Z'\left(\frac{\hbar \omega}{k_B\theta}\right),
\nonumber \\
\sum_{r=0}^n \binom{n}{r} \exp\left(- \frac{r \hbar \omega}{k_B \theta}\right) r^2 & = \left(\frac{k_B \theta^2}{\hbar \omega}\right)^2 \frac{\lbrace Z'[\hbar \omega /(k_B\theta)]\rbrace^2}{Z[\hbar \omega /(k_B\theta)]}\left\lbrace1 + \frac{\exp\left[\hbar \omega/(k_B \theta)\right]}{n} \right\rbrace,
\label{fermionIdentities}
\end{align}
with 
\begin{equation}
Z'\left(\frac{\hbar \omega}{k_B\theta}\right) = \frac{n \hbar \omega}{k_B \theta^2} \frac{Z[\hbar \omega /(k_B\theta)]}{\exp[\hbar \omega/(k_B \theta)] + 1},
\end{equation}
which can be found by differentiating 
\begin{equation}
\sum_{r=0}^n \binom{n}{r} \exp\left(- \frac{r \hbar \omega}{k_B \theta}\right)    
\end{equation}
with respect to $\theta$ once and twice, respectively. Then, inserting Eq. (\ref{fisherSupp}) into Eq. (\ref{asymSupp}), for $k_B \theta/(\hbar \omega) \in [0.1, 10]$, leads to 
\begin{equation}
\bar{\epsilon}_{\mathrm{cr}} = \frac{2}{n \log(10)} \left(\frac{k_B}{\hbar \omega} \right)^2 \int_{0.1 \hbar \omega/k_B}^{10 \hbar \omega/k_B} d\theta\,\theta\,\cosh^2\left(\frac{\hbar \omega}{2 k_B \theta}\right)
= \frac{2}{n \log(10)} \int_{0.1}^{10} dy\,y\,\cosh^2\left(\frac{1}{2y}\right)
\simeq \frac{51.7}{n},
\end{equation}
where $k_B \theta/(\hbar \omega) \coloneqq y$ and the value for the last integral has been found numerically. In summary, we have shown that, asymptotically, $\bar{\epsilon}_{\mathrm{opt}} \simeq \bar{\epsilon}_{\mathrm{cr}} = 51.7/n$, as stated in the main discussion. 

\subsection{Data analysis in global thermometry}

\subsubsection{Protocol}

Consider the measurement protocol:
\begin{enumerate}
    \item a gas of $n$ non-interacting spin-$1/2$ particles is prepared with statistics according to $p(r|\theta)$ in Eq.~\eqref{likelihoodFermionsSupp};
    \item its total energy $r\hbar \omega$ is measured; and
    \item both steps are repeated $\mu$ times, generating the outcomes $(r_1,\dots,r_{\mu})\coloneqq \boldsymbol{r}$.
\end{enumerate}
The likelihood function representing the information associated with this protocol is
\begin{equation}
    p(\boldsymbol{r}|\theta) = \prod_{i=1}^{\mu} p(r_i|\theta) = \left[ \prod_{i=1}^\mu \binom{n}{r_i} \right] 
    \exp\left(- \frac{\mu \bar{r} \hbar \omega }{k_B \theta}\right)
    \left[\exp\left(- \frac{\hbar \omega }{k_B \theta}\right) + 1 \right]^{-\mu n},
    \label{likelihood_multishot_protocol}
\end{equation}
where $\bar{r} \coloneqq \sum_{i=1}^{\mu} r_i/\mu$.

\subsubsection{Simulation of outcomes}

For the purpose of illustrating \emph{how} an analysis of experimental data would proceed using the global framework in this work, here we simulate the measurement outcomes $(r_1,\dots,r_{\mu})$. First we construct the discrete cumulative distribution 
\begin{equation}
     \sum_{m=0}^r p(m|T) \coloneqq f_{n,T}(r),
\end{equation}
where $p(m|T)$ is given by Eq.~\eqref{likelihoodFermionsSupp}, $T$ is the true temperature, and $0 \leqslant r \leqslant n$. By construction, $ 0 \leqslant f_{n,T}(r) \leqslant 1 $. Next, we impose $f_{n,T}(r) = u$, where $u$ is a uniformly distributed random number between $0$ and $1$. Inverting this relationship we get 
\begin{equation}
    r = f_{n,T}^{-1}(u),
\end{equation}
where $f_{n,T}^{-1}$ is defined such that $f_{n,T}^{-1}(f_{n,T}(x)) = x$ and $f_{n,T}(f_{n,T}^{-1}(y)) = y$. Finally, by generating a string of $\mu$ random numbers $(u_1,\dots, u_{\mu})$ between $0$ and $1$, we arrive at
\begin{equation}
    (r_1,\dots,r_{\mu}) = (f_{n,T}^{-1}(u_1), \dots, f_{n,T}^{-1}(u_{\mu})).
\end{equation}
The data in the main text was simulated with a numerical implementation of this procedure. We chose $k_B T/(\hbar \omega) = 4 $, $n = 150$, and $\mu = 500$, and the random numbers were produced using the \texttt{rand} function in MATLAB. 

\subsubsection{Global thermometry in experiments}

Consider the logarithmic deviation function $\mathcal{D}[\tilde{\theta}(\boldsymbol{r}),\theta] = \log^2[\tilde{\theta}(\boldsymbol{r})/\theta]$. To the theorist who is searching for fundamental precision limits, both the correct hypothesis $\theta$ and the measurement outcomes $\boldsymbol{r}$ that a given scheme may produce are unknown. Hence, theoretical studies such as that in the first part of our work require that we average $\log^2[\tilde{\theta}(\boldsymbol{r})/\theta]$ over both temperatures and outcomes weighted by the joint probability $p(\boldsymbol{r},\theta)$, which leads to the uncertainty quantifier \footnote{Note that here we have already adapted the generic notation in the main text to the specific spin system under consideration.}
\begin{equation}
    \bar{\epsilon}_{\mathrm{mle}} = \sum_{\boldsymbol{r}}\int d\theta\,p(\boldsymbol{r},\theta)\log^2\left[\frac{\tilde{\theta}(\boldsymbol{r})}{\theta} \right].
    \label{errorMLEapp}
\end{equation}
The situation an experimentalist faces is, however, different. Since, in experiments, the measurement outcomes $\boldsymbol{r}$ are known (here, because we simulated them), the uncertainty quantifier for this new situation is instead
\begin{equation}
    \bar{\epsilon}_{\mathrm{mle}}(\boldsymbol{r}) \coloneqq \int d\theta\,p(\theta|\boldsymbol{r})\log^2\left[\frac{\tilde{\theta}(\boldsymbol{r})}{\theta} \right],
    \label{errorMLEappExp}
\end{equation}
where $p(\theta|\boldsymbol{r})$ is the posterior probability given, in this case, by 
\begin{equation}
    p(\theta|\boldsymbol{r}) \propto p(\theta) p(\boldsymbol{r}|\theta) \propto 
    \frac{1}{\theta}
     \exp\left(- \frac{\mu \bar{r} \hbar \omega }{k_B \theta}\right)
    \left[\exp\left(- \frac{\hbar \omega }{k_B \theta}\right) + 1 \right]^{-\mu n},
\end{equation}
with $k_B \theta/(\hbar \omega) \in [0.1, 10]$. As expected, the temperature is still integrated---it is unknown---, but the final error depends now on $\boldsymbol{r}$. Note that the deviation function is still $\mathcal{D}[\tilde{\theta}(\boldsymbol{r}),\theta] = \log^2[\tilde{\theta}(\boldsymbol{r})/\theta]$.

A natural question is whether the estimator in Eq.~\eqref{optEstSupp} is optimal also with respect to the uncertainty quantifier in Eq.~\eqref{errorMLEappExp}. The next section answers in the affirmative---even when theorists and experimentalists need to evaluate their uncertainties differently because their initial information differs, both use the same deviation function and will find the same optimal estimator. For an extended discussion, see, e.g., sections 13.8 and 13.8 of Ref.~\cite{jaynes2003}, and section 3.2 of Ref.~\cite{jesus2019thesis}.

\subsubsection{Optimal strategy}

The first derivative of Eq.~\eqref{errorMLEappExp} reads
\begin{align}
    \frac{d \bar{\epsilon}_{\mathrm{mle}}(\boldsymbol{r})}{d \tilde{\theta}(\boldsymbol{r})} = 2 \int d\theta\,\frac{p(\theta|\boldsymbol{r})}{\tilde{\theta}(\boldsymbol{r})} \left\lbrace \log\left[\frac{k_B \tilde{\theta}(\boldsymbol{r})}{\varepsilon_0}\right] - \log\left(\frac{k_B \theta}{\varepsilon_0} \right) \right\rbrace.
\end{align}
Imposing $d \bar{\epsilon}_{\mathrm{mle}}(\boldsymbol{r})/d \tilde{\theta}(\boldsymbol{r}) = 0$ leads to 
\begin{equation}
\frac{k_B \tilde{\theta}(\boldsymbol{r})}{\varepsilon_0} 
= \exp\left[\int d\theta\,p(\theta|\boldsymbol{r})\log\left(\frac{k_B \theta}{\varepsilon_0}\right) \right] \coloneqq \frac{k_B \tilde{\vartheta}(\boldsymbol{r})}{\varepsilon_0},
\label{optEstSuppSpin}
\end{equation}
which is the same estimator in Eq.~\eqref{optEstSupp} but written in the notation of the spin gas. By inserting Eq.~\eqref{optEstSuppSpin} in the expression for the second derivative $d^2 \bar{\epsilon}_{\mathrm{mle}}(\boldsymbol{r})/d \tilde{\theta}(\boldsymbol{r})^2$, one further finds that the latter is positive. Therefore, the estimator in Eq.~\eqref{optEstSuppSpin} (or, equivalently, Eq.~\eqref{optEstSupp}) leads to a minimal error and is thus optimal with respect to both the theory-motivated uncertainty in Eq.~\eqref{errorMLEapp}, and the experiment-based uncertainty in Eq.~\eqref{errorMLEappExp}. 

\subsection{Data analysis in local thermometry}

In the absence of an asymptotically large number of measurement outcomes (i.e., $\mu$ is finite and possibly small), local estimation leads to estimates of the form \cite{kay1993, rafal2015}
\begin{equation}
    \tilde{\theta}_L(\boldsymbol{r}) \pm \Delta \tilde{\theta}_L = \left\lbrace\theta_0 + \frac{1}{\mu F(\theta_0)} \frac{\partial \log[p(\boldsymbol{r}|\theta)]}{\partial \theta}\Bigg\vert_{\theta=\theta_0}\right\rbrace \pm \frac{1}{\sqrt{\mu F(\theta_0)}},
    \label{localApproachSupp}
\end{equation}
where $\theta_0$ is an initial `hint' at the true temperature. 

From equation Eq.~\eqref{likelihood_multishot_protocol} we see that 
\begin{equation}
    \frac{\partial \log[p(\boldsymbol{r}|\theta)]}{\partial \theta} = \frac{\mu \hbar \omega}{k_B \theta^2}\left\lbrace \bar{r} -
    \frac{n}{\exp[\hbar \omega/(k_B \theta)]+1} \right\rbrace.
\end{equation}
By inserting this and the Fisher information \eqref{fisherSupp} in Eq.~\eqref{localApproachSupp}, we find that the local estimate associated with the $n$-spin gas is
\begin{equation}
    \tilde{\theta}_L(\boldsymbol{r}) = \theta_0 \left\lbrace 1 + \frac{k_B \theta_0}{\hbar \omega} \left[ \frac{4 \bar{r} }{n} \cosh^2\left(\frac{\hbar\omega}{2 k_B \theta_0} \right) - \exp\left(- \frac{\hbar \omega}{k_B \theta_0}\right) -1 \right] \right\rbrace
\end{equation}
with uncertainty
\begin{equation}
    \Delta \tilde{\theta}_L = \frac{1}{\sqrt{\mu n}}\frac{2 k_B \theta_0^2}{\hbar \omega} \cosh\left(\frac{\hbar\omega}{2 k_B \theta_0} \right). 
\end{equation}
The local estimator examined in the main text is based on the choice $k_B \theta_0/(\hbar \omega) = 3$, which is close to the true temperature $k_B T/(\hbar \omega) = 4$. All other simulation parameters were chosen as in indicated in the previous section.

\subsection{Thermometry beyond energy measurements}

Scale-invariant global thermometry is not restricted to protocols that measure energy. For example, consider a harmonic oscillator with mass $m$, angular frequency $\omega$ and coordinate $q$. Measuring the position, the probability that the dimensionless position coordinate $x\coloneqq q \sqrt{m\omega/\hbar}$ lies between $x$ and $x+dx$ is given, in the canonical ensemble, by \cite{pathria2011}
\begin{equation}
p(x|\theta)dx = \frac{\mathrm{exp}\lbrace - x^2/[2\,\sigma(\theta)^2] \rbrace}{\sqrt{2\pi \sigma(\theta)^2}}\,
dx,
\label{eq:positionEx}
\end{equation}
where 
\begin{equation}
    \sigma(\theta) = \sqrt{\frac{1}{2} \coth\left(\frac{\hbar \omega}{2 k_B \theta}\right)}.
    \label{eq:positionExdetail}
\end{equation}
We see that, even when energy is no longer the measured quantity, Eq.~\eqref{eq:positionEx} only depends on the ratio of energies $\hbar \omega/2$ and $k_B \theta$. As such, the arguments leading to the prior $p(\theta) \propto 1/\theta$ and the deviation function $\mathcal{D}[\tilde{\theta}(\boldsymbol{x}),\theta] = \log^2[\tilde{\theta}(\boldsymbol{x})/\theta]$ apply.

\begin{figure}[t]
\centering
\includegraphics[trim={0cm 0.75cm 1cm 0.5cm},clip,width=9cm]{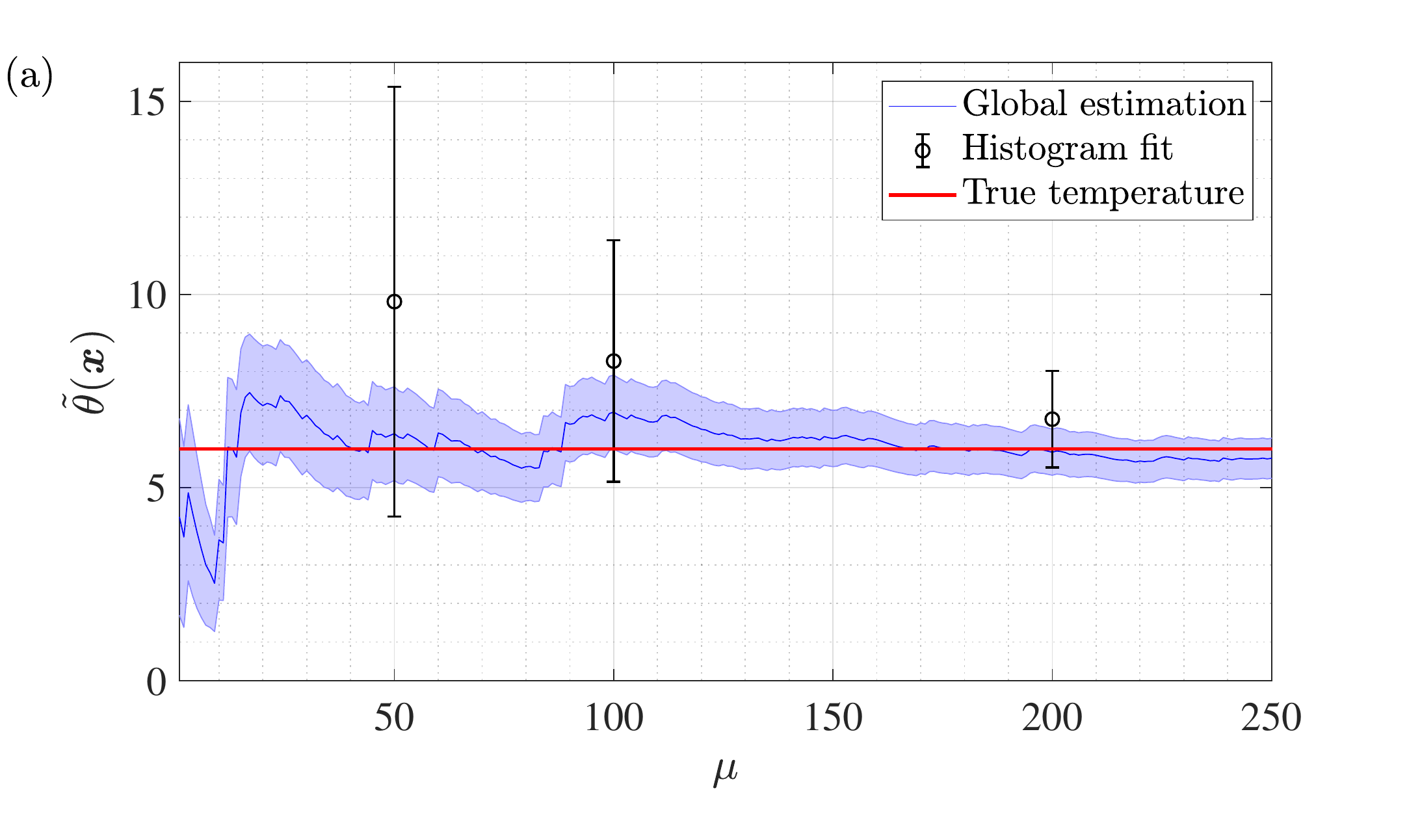}\includegraphics[trim={0cm 0.75cm 1cm 0.5cm},clip,width=9cm]{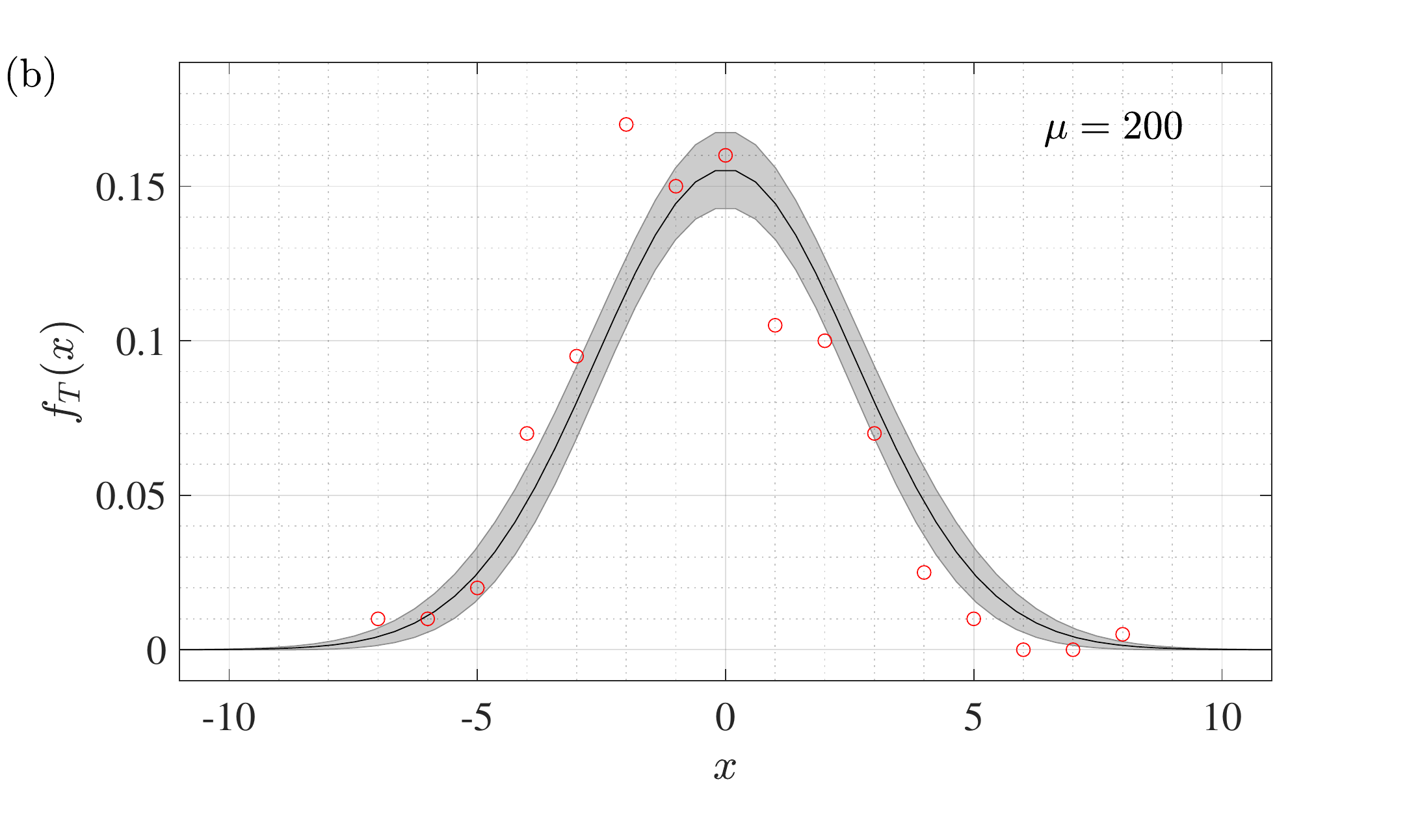}
\includegraphics[trim={0cm 0.75cm 1cm 0.5cm},clip,width=9cm]{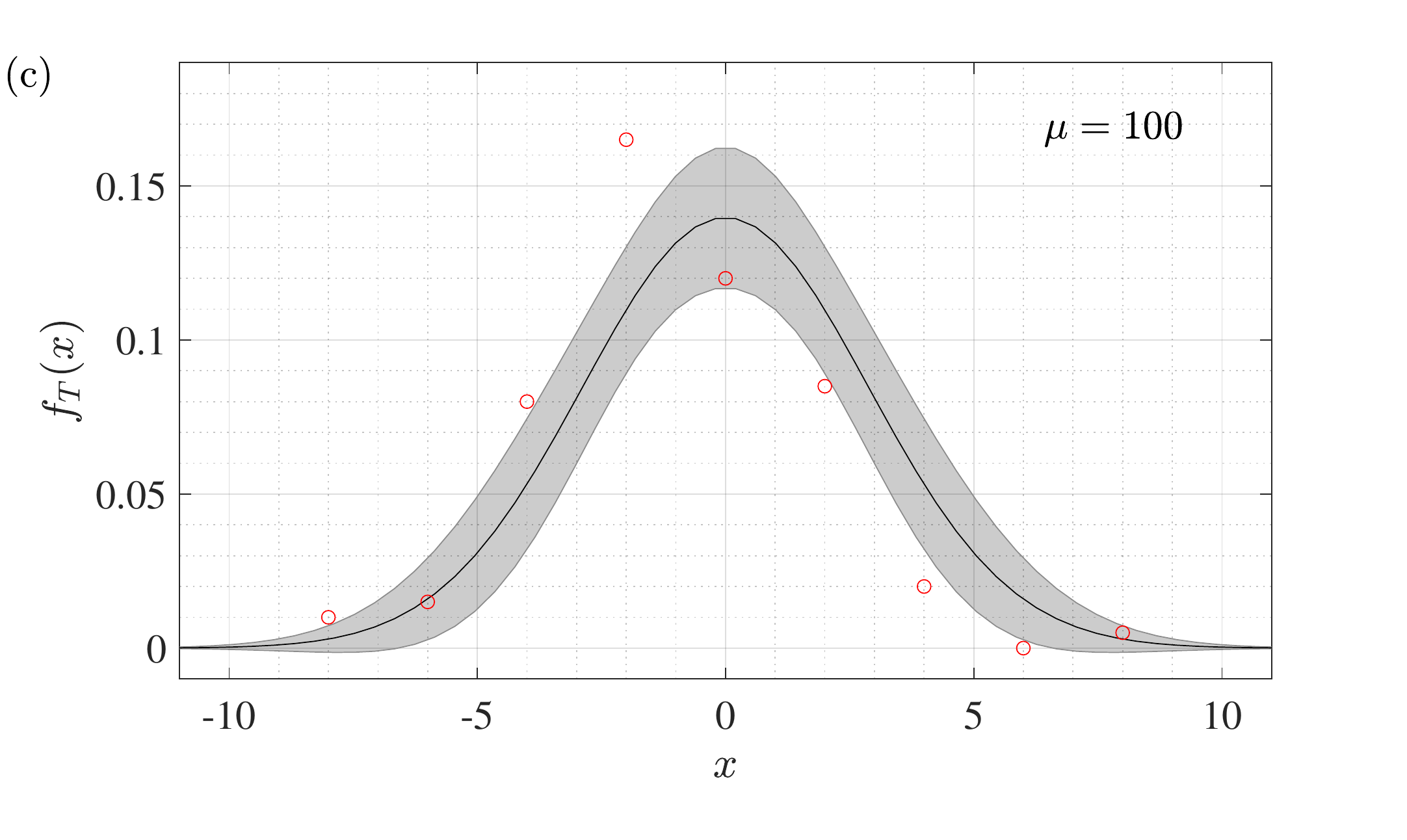}\includegraphics[trim={0cm 0.75cm 1cm 0.5cm},clip,width=9cm]{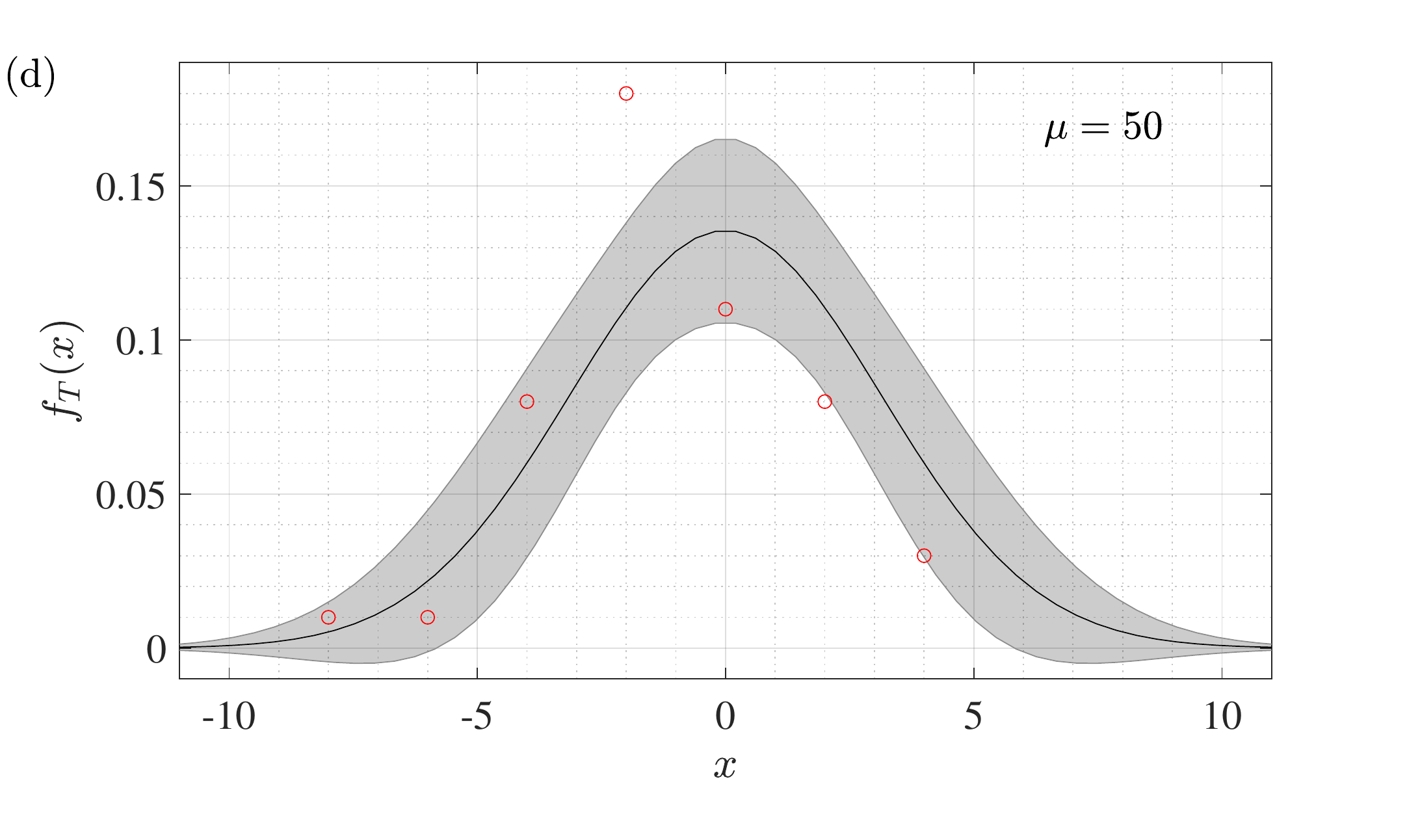}
	\caption{(colour online) a) Global estimate (shaded blue) from position measurements $\boldsymbol{x}$ on a quantum harmonic oscillator in the canonical ensemble. This is to be compared with the fit of position histograms to a Gaussian profile $A\, \mathrm{exp}[-x^2/(2\sigma^2)]$, from where three temperature estimates (black circles), for $\mu = 50$, $100$ and $200$, have ben generated. Specifically, these are found by inverting the relationship between variance $\sigma$ and temperature $\theta$ in Eq.~\eqref{eq:positionExdetail}. As can be observed, the global estimates oscillate around the true temperature (solid red) even when the number of trials $\mu$ is low. In contrast, the asymptotic estimates are biased towards larger temperatures unless $\mu \gtrsim 200$. Plots (b-d) show the aforementioned histograms (red circles) together with the associated Gaussian fits (solid black). The uncertainty areas (shaded black) have been calculated by propagating the standard errors of the estimates for $A$ and $\sigma$.}
\label{positionBayes}
\end{figure}

To verify that global thermometry renders the correct temperature in this scenario, we simulated a string of outcomes $\boldsymbol{x} = (x_1, \dots, x_\mu)$ drawn from the thermal distribution in Eq.~\eqref{eq:positionEx}, with  true temperature $k_B T/(\hbar \omega) = 6$. Given the posterior probability $p(\theta|\boldsymbol{x})\propto p(\theta)\,\Pi_{i=1}^\mu p(x_i|\theta)$, with $k_B\theta/(\hbar \omega) \in [0.1, 10]$, the calculation of the global estimate $\tilde{\vartheta}(\boldsymbol{x})\pm \Delta \tilde{\vartheta}(\boldsymbol{x})= \tilde{\vartheta}(\boldsymbol{x})[1 \pm \sqrt{\bar{\epsilon}_{\mathrm{mle}}(\boldsymbol{x})}]$ proceeds as in the main text. The result, represented as a blue shaded area in Fig.~\ref{positionBayes}~a), shows how the global estimate oscillates around the correct temperature (red solid line) and converges from $\mu \sim 100$ trials.

\subsubsection{Global estimation vs histogram fitting}

To provide further motivation for the application of global techniques in experiments, let us compare the new global approach with the standard practice of fitting probability distributions, which is an asymptotic estimation method. Let $f_T(x)dx$ be the relative frequency of the dimensionless coordinate lying between $x$ and $x + dx$ when the true temperature is $T$. If $\mu \gg 1$, it is often reasonable to assume that $f_T(x)dx \sim p(x|\theta = T)dx$. In turn, this justifies fitting a histogram constructed with the outcomes $\boldsymbol{x}$ to the Gaussian profile in Eq.~\eqref{eq:positionEx}. Such a procedure yields an estimate $\sigma \pm \Delta \sigma$ for the variance in Eq.~\eqref{eq:positionExdetail}; we can then estimate the temperature by inverting the functional relationship between $\sigma$ and $\theta$, finding $k_B\tilde{\theta}_F(\boldsymbol{x})/(\hbar \omega) = 1/[2\,\mathrm{arcoth} (2 \sigma^2)]$. The associated uncertainty $\Delta \tilde{\theta}_F(\boldsymbol{x})$ can further be calculated by propagating the standard error $\Delta \sigma$.

Considering the \emph{same} dataset $\boldsymbol{x}$ employed in the Bayesian analysis (Fig.~\ref{positionBayes}~a), blue), we constructed the histograms shown in Figs.~\ref{positionBayes}~b) - d) for the first $\mu = 200$, $100$ and $50$ data points, respectively. The asymptotic estimates are $k_B (\tilde{\theta}_F \pm \Delta \tilde{\theta}_F)/(\hbar \omega) = 7 \pm 1$, $8 \pm 3$ and $10 \pm 6$, and these are shown in Fig.~\ref{positionBayes}~a) as black circles with error bars. While these asymptotic estimates correctly converge to the true temperature as the number of data points increases, they are noticeably biased towards large temperatures when the number of trials $\mu$ is small. In contrast, not only are the global counterparts (blue) in Fig.~\ref{positionBayes}~a) closer to the correct temperature even when the number of trials is as low as $\mu = 50$, but they also include values both above and under the true temperature, as one would expect from a reliable limited-data estimate. 

This example, together with the local analysis for a gas of spin-$1/2$ particles considered in previous sections, constitutes a solid demonstration that global thermometry can have an appreciably better performance than standard techniques. Moreover, we note that global techniques bypass the necessity of constructing histograms, thus holding great potential for applications where the number of trials is small. Importantly, spatially resolved position measurements of impurity atoms thermalising with a majority condensate have been experimentally demonstrated by using an optical lattice to `freeze' the impurities in place with micrometre resolution prior to fluorescence imaging \cite{PhysRevA.93.043607}.

\end{document}